\documentclass[aps,pre,superscriptaddress,twocolumn,showpacs,floatfix]{revtex4}
\usepackage{graphicx}% Include figure files
\usepackage{dcolumn}% Align table columns on decimal point
\usepackage{bm}% bold math
\newcommand{\lan}{\langle}
\newcommand{\ran}{\rangle}

\begin{document}
\title{Aging and species abundance in the Tangled Nature model of biological evolution. }
\author{Matt Hall} 
\affiliation{Department of Mathematics, Imperial College, 
180 Queen's Gate, London SW7 2BZ, U.K.}
\author{Kim Christensen}
\author{Simone A. di Collobiano}
\affiliation{Blackett Laboratory, Imperial College, Prince Consort Road, London SW7 2BW,
U.K. }
\author{ Henrik J. Jensen}
\email[Author to whom correspondence should be addressed:\\]
{h.jensen@ic.ac.uk} 
\homepage{http://www.ma.ic.ac.uk/~hjjens/}
\affiliation{Department of Mathematics, Imperial College, 
180 Queen's Gate, London SW7 2BZ, U.K.}

\date{4. Feb. 02}

\begin{abstract}
% insert abstract here
We present an individual based model of evolutionary ecology.
 The reproduction rate of individuals characterized by their 
 genome depends on the composition of the population in genotype 
 space. Ecological features such as the taxonomy and the 
 macro-evolutionary mode of the dynamics are emergent properties. 
 The macro-dynamics exhibit intermittent two mode switching with a 
 gradually decreasing extinction rate. The generated ecologies become 
 gradually better adapted as well as more complex in a collective sense. 
 The form of the species abundance curve compares well with observed 
 functional forms. The models error threshold can be understood in 
 terms of the characteristics of the two dynamical modes of the system.

\end{abstract}

\maketitle

\section{Introduction}
The dynamics and organization of biological ecosystems is a fascinating example of complex interacting systems with many levels of emerging structure and time scales. Biological evolution creates intricate taxonomic hierarchies presumably as an effect of mutation, natural selection and the ensuing adaptation. Taxonomic structures from the level of individuals through species and genera up to kingdoms are generated and
vanish again in a never ending succession. Different strata in the hierarchy are described by very different timescales and with very different types of dynamics. At the level of individuals, fairly well defined characteristic lifetimes exist for each specific type (species) and the population dynamics can be considered smooth. 
This picture changes as one considers the system at the more coarse grained level of species and genera. The lifetime distribution of, e.g., genera is broad (see e.g. \cite{newman}) and the dynamics is intermittent \cite{eldredge,gould,gould_eldredge,eldredge_gould}.
In the spirit of the traditional approach of statistical mechanics it is interesting to consider models, defined at a microscopic level, which are able to reproduce the large scale temporal and taxonomic structures.

In the present paper we consider a model of individuals identified solely by their genome. The model was introduced in Ref. \cite{theo_bio_p} where we also presented a discussion of the qualitative behavior of the model.  We combine ecology with evolution by considering interacting individuals which can multiply (sexually or asexually) subject, potentially, to mutations.  The size of the total population fluctuates, the average being controlled by the amount of available resources. From these three minimal ingredients emerge segregation in genome space, to be interpreted as the appearance of species, and a complex intermittent dynamics, to be interpreted as extinction and creation events at the higher taxonomic levels. The entire taxonomic hierarchy is an emergent property of the dynamics at the microscopic level of individuals. We characterize the configurations generated in genotype space in terms of the species abundance curve, and find a good qualitative agreement with the functional form typically found for real ecosystems. The intermittent dynamics is characterized by the statistics of the duration of the quasi-stable epochs or in other words the waiting times between transitions. We find a broad distribution of durations and observe a gradual aging of the macro-dynamics. No stationary state is ever reached.

\subsection{Related Models}
Many mathematical models of biological evolution have been developed according to the usual statistical mechanics agenda of generating the macroscopic complex behavior from simplistic microscopic definitions. An elegant review of this endeavor has recently been given by Drossel \cite{drossel_rev}. Here we limit ourselves to a discussion of similarities and differences between our model and related studies.

Let us first mention models which define the ecosystem in terms of individuals. Higgs and Derrida \cite{higgs1} studied speciation in a model consisting of a fixed number of individuals. Each individual is represented by a genome modeled as a string of zeros or ones, like in Eigen and collaborators' seminal work on quasi-species \cite{eigen}. Higgs and Derrida demonstrated that a sexually reproducing population breaks up into distinct species when only individuals with a sufficiently similar genome sequence are allowed to produce offspring. This agrees with a large bulk of experimental work \cite{hostert}.  Gavrilets and collaborators \cite{gavrilets,glv_1,glv_2} have made use of similar models generalized in particular to be able to study geographical and temporal aspects of speciation. These studies differ from ours in assuming a fixed population size and by defining a fitness function for pairs of individuals which is constant if the Hamming distance between the genomes of two individuals is small enough, and zero otherwise. Our model allows the total size of the population to fluctuate and the fitness of pairs of individuals (or in the asexual case single individuals) depends on the composition of the population at a given instant in time. 

It is also important to mention the fitness landscape approach first pioneered by Wright \cite{wright_1,wright_2}, who considered gene frequencies, and was brought to the attention of the statistical mechanics community mainly through Kauffman's so called NK model \cite{NK-model,kauffman}. The main focus of the NK model, and of the later co-evolutionary NKC model \cite{kauffman}, is the study of epistatic interactions (the influence of one gene on another) by use of fitness functions. The main difference between our model and Kauffman's models is that the fitness of an individual in our system depends on the {\em frequencies} by which other locations in genotype space are occupied.

Taylor and Higgs \cite{taylor} have studied pleiotropy and epistasis (the influence of one gene on several traits and the influence of one gene on another) in a model that combines and generalizes aspects of the Higgs-Derrida model with the epistatic interactions of Kauffman's models. Taylor and Higgs then derive a phenotypical fitness for the specific genotype. Kaneko and Yomo \cite{kaneko} have also studied models in which the difference between phenotype and genotype is accounted for explicitly. In our model we make the drastic simplification not to distinguish between genotype and phenotype.

Other models consider species as the elementary building block; these models neglect the specifics of the dynamics arising from reproduction and mutations at the level of individuals. The simplest of these models is the Bak-Sneppen model \cite{bak_sneppen}. The model aims to demonstrate that co-evolutionary interactions are sufficient to produce intermittent dynamics which is then related to intermittency in the fossil record and to Eldredge and Gould's concept of punctuated equilibrium 
\cite{eldredge,gould,gould_eldredge,eldredge_gould}. Each species is characterized by a single number between zero and one, the fitness, and the total number of species is kept constant. The model has interesting statistical properties but is difficult to relate to biological evolution.

Species level models of more detail than the Bak-Sneppen model  have been formulated recently by McKane, Alonso and Sol\'e \cite{mckane} and by Drossel,
Higgs and McKane \cite{drossel_2}. The emphasis in these models is on predator-prey interactions and food-webs and are generalizations of early work by May \cite{may} and May and Anderson \cite{may_anderson}. Our model is intended to include all types of interactions between individuals, e.g. antagonistic or collaborative relationships, in addition to predator-prey competitions. Another important difference is that we define our model at the level of individuals in order to be able to study the emergence of species, something not possible in a species based model.

Most models of biological evolution assume that the dynamics is in a statistically stationary state. One marked exception is the model considered by Sibani and collaborators \cite{sibani_1,sibani_2,sibani_3,sibani_4}. This is an abstract species based model consisting of random walks in a rugged fitness landscape. The statistics of the jumps in this landscape are the same as the record statistics considered some time ago by Sibani and Littlewood \cite{sibani_littlewood}. The pace of the dynamics of the model gradually slows down as indicated by a logarithmically decreasing extinction rate. As we shall see below our individual based model also exhibits aging, a property found to be consistent with analysis of the fossil record \cite{newman}.

The paper is organized as follows. In the next section we define the model in detail.
In Sec. \ref{dyn_stab} we discuss the modes of the models emergent dynamics. In Sec. \ref{aging} we show how the configurations generated dynamically gradually become better adapted in a collective sense. Sec. \ref{spec_ab} demonstrates that the ecologies generated in the model exhibit characteristics similar to those observed in real ecologies. In Sec. \ref{competition} we discuss the models ability to address the issue concerning why sexual reproduction can compete evolutionary with asexual reproduction. Sec. \ref{error} contains an analysis of the error threshold. We briefly present in Sec. \ref{para} a scan of the behavior of the model for a range of the control parameters and in Sec. \ref{conclusion} we conclude and summarize. 

\section{Definition of Model}
We describe here in detail the structure and dynamics of the model which we, with an allusion to Darwin's notion of the {\it Tangled Bank}, called the Tangled Nature (TaNa) model to stress the model's emphasis on ecological interactions  
 \cite{theo_bio_p}.

\subsection{Interaction}
We represent an individual by a vector ${\bf S}^\alpha=
(S_1^\alpha,S_2^\alpha,...,S_L^\alpha)$ in genotype space $\cal{S}$.
This representation is frequently used, see e.g.  Ref. \cite{eigen,kauffman,higgs1,gavrilets,wagner}. Here $S^\alpha_i$ may take the values $\pm 1$, i.e. ${\bf S}^\alpha$ denotes one of the corners of the $L$ dimensional hypercube (in the present paper we use $L=20$). The coordinates $S^\alpha_i$ may be interpreted as genes with two alleles, or a string of either pyrimidines or purines. We think of genotype space $\cal{S}$ as containing all possible ways of combining the genomic building blocks into genome sequences. Many sequences may not correspond to viable organisms. Whether this is the case or not is for the evolutionary dynamics to determine. All possible sequences are made available for evolution to select from.

Individuals are labeled by Greek letters  $\alpha,\beta, ... = 1,2,..., N(t)$.  When we refer, without reference to a specific individual, to one of the $2^L$ positions in genome space, we use roman superscripts ${\bf S}^a$, ${\bf S}^b$, ... with  $a,b,... =1,2,...,2^L$. Many different individuals  ${\bf S}^\alpha, {\bf S}^\beta$,..., may reside on the same position, say ${\bf S}^a$, in $\cal{S}$. 

The ability of an individual $\alpha$ to reproduce is controlled by $H({\bf S}^\alpha,t)$:  
\begin{equation}
H({\bf S}^\alpha,t)={1\over c N(t)}\sum_{{\bf S}\in{\cal S}} 
J({\bf S}^\alpha,{\bf S})   n({\bf S},t)
- \mu N(t),
\label{Hamilton2} 
\end{equation}
where $c$ is a control parameter (see below), $N(t)$ is  the total number of individuals at time $t$, the sum is over the $2^L$ locations ${\bf S}$ in ${\cal S}$ and $n({\bf S},t)$ is the occupancy of position ${\bf S}$. Two positions ${\bf S}^a$ and ${\bf S}^b$ in genome space are coupled with the fixed random strength $J^{ab}=J({\bf S}^a,{\bf S}^b)$ which can be either positive or negative or zero. The coupling is non-zero with probability $\Theta$ (throughout the paper we use $\Theta=0.25$), in which case we assume $J^{ab}\neq J^{ba}$ to be a deterministic but erratic function of the two positions ${\bf S}^a$ and ${\bf S}^b$. We have checked that the specific details of the form of the distribution of the non-zero values of the function  $J({\bf S}^a,{\bf S}^b)$ are irrelevant. We choose accordingly a form mainly determined by its numerical efficiency.
In the next subsection we describe the details of the specific procedure used. The distribution of the generated interaction strengths is shown in Fig. \ref{P(J)} below.

\subsubsection{Generation of interaction matrix}
The interaction between two locations in genotype space, ${\bf S}^a$ and ${\bf S}^b$ is generated 
as a product $J({\bf S}^a,{\bf S}^b)= \Theta({\bf S}^a,{\bf S}^b) I({\bf S}^a,{\bf S}^b)$. The first factor $\Theta({\bf S}^a,{\bf S}^b)$ is obtained by interpreting the sequences ${\bf S}^a$ and ${\bf S}^b$ as binary numbers (letting 
$-1\mapsto 0$) and perform the \textsf{XOR} operation on the binary pair to obtain a new integer. This integer is used as an index in a lookup list to obtain either a 0 or 1 as the value of $\Theta({\bf S}^a,{\bf S}^b)$ . In case 1 is returned, the element of the $I({\bf S}^a,{\bf S}^b)$ matrix is obtained in a similar way. This time, however, two arrays are needed. Each auxiliary array is of length $2^L$ and now the arrays contain uniformly distributed random numbers drawn from the interval $[-1,+1]$. The pair of arrays is necessary in order to reproduce the asymmetry of the $I({\bf S}^a,{\bf S}^b)$ matrix. Two indices are generated from the ${\bf S}^a$ and
${\bf S}^b$. The first via the same \textsf{XOR} operation used to calculate the $\Theta({\bf S}^a,{\bf S}^b)$ matrix element, whereas the second is simply the integer representing ${\bf S}^b$. The strength of interaction is taken to be the product of the members of each array at the appropriate location. This ensures that the elements of the matrices are non-symmetric due to the second array index depending on the order of the operation. This procedure is numerically extremely efficient and deterministic, but has the side effect of generating a distribution of a slightly unusual form, see Fig. \ref{P(J)}.

We stress that the coupling matrix $J({\bf S}^a,{\bf S}^b)$  is meant to included {\em all} possible interactions between two individuals of a given genomic constitution. In our simplistic approach, a given genome is imagined to lead uniquely to a certain set of attributes (phenotype) of the individuals/organisms. The locations ${\bf S}^a$ and ${\bf S}^b$ represent blueprints for organisms that exist {\it in potentia}. The positions may very likely be unoccupied but, if we were to construct individuals according to the sequences ${\bf S}^a$ and ${\bf S}^b$ the two individuals would have some specific features. The relationship between an organism of {\it design} ${\bf S}^a$ and one of design ${\bf S}^b$ may be as predator and prey or parasitic, i.e. $J^{ab}>0$ and $J^{ba}<0$, but it can also be collaborative ($J^{ab}>0$ and $J^{ba}>0$) or antagonistic ($J^{ab}<0$ and $J^{ba}<0$), see Fig. \ref{J^{ab}}.  And certainly in some cases $J^{ab}$ may represent less direct couplings, e.g. some animals may not eat trees, nevertheless they breath the oxygen produced by the rain forest. In order to emphasis co-evolutionary aspects we have {\em excluded} ``self-interaction'' among individuals located at the same position  ${\bf S}$ in genome space, i.e. $J({\bf S},{\bf S})={\bf 0}$ for all  $\bf S \in {\cal S}$. It is important to mention that including self-interactions of the same
 order of strength as the $J$-couplings does not change the qualitative behavior of the model.

\begin{figure}
\includegraphics*[scale=.7]{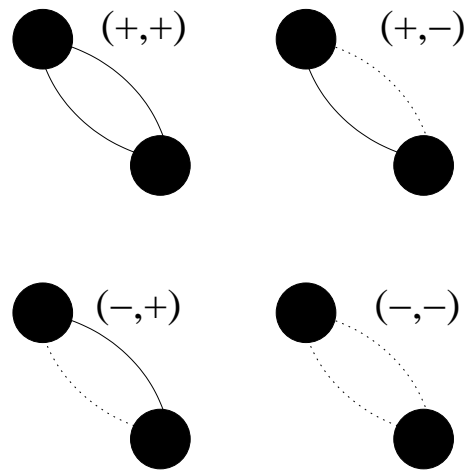}
\caption{\label{J^{ab}} Examples of possible realizations
of the couplings $J^{ab}$ between different positions
${\bf S}^a$ and ${\bf S}^b$ in genotype space representing collaborative (+,+),
antagonistic (-,-) or predator prey (+,-) \& (-,+) relationships.
} 
\end{figure}

The conditions of the physical environment are simplistically described by the term $\mu N(t)$ in Eq. (1), where $\mu$ determines the average sustainable total population size. That is, the total carrying capacity of the environment. An increase in $\mu$ corresponds to harsher physical conditions. This is a simplification, though one should remember that what is often considered as the physical conditions, e.g. temperature or oxygen density, is to a degree determined by the activity of other organisms and is therefore really a part of the biotic conditions. Consider, for example, the environment experienced by the bacterial flora in the intestines. Here one type of bacteria live very much in an environment strongly influenced by the presence of other types of bacteria. In this sense some fluctuations in the environment may be thought of as included in the coupling matrix $J({\bf S}^a,{\bf S}^b)$.

\subsection{\label{reproduction} Reproduction, mutations and annihilation}
Asexual reproduction consists of one individual being replaced by two copies.
Successful reproduction occurs for individuals ${\bf S}^\alpha$ with a probability per time unit given by
\begin{equation}
p_{off}({\bf S}^\alpha,t)={ \exp[H({\bf S}^\alpha,t)]\over
1+\exp[H({\bf S}^\alpha,t)]}\in[0,1].
\label{p_off}
\end{equation}

In the case of sexual reproduction an individual ${\bf S}^\alpha$ is picked at random and paired with another randomly chosen individual ${\bf S}^\beta$ with Hamming distance $d={1\over2}\sum_{i=1}^L|S_i^\alpha-S_i^\beta|\leq d_{max}$ (allowing at most $d_{max}$ pairs of genes to differ). The pair produces an offspring
$\gamma$ with probability $\sqrt{p_{off}({\bf S}^\alpha,t)p_{off}({\bf S}^\beta,t)}$, where $S_i^\gamma$ is chosen at random from one of the two parent genes, either $S_i^\alpha$ or $S_i^\beta$. For $d_{max}\geq1$ this procedure may be thought of as being similar to recombination. The maximum separation criterion has been studied by several authors, see e.g. \cite{higgs1,gavrilets}. 

We allow for mutations in the following way: with probability $p_{mut}$ per gene we perform a change of sign $S_i^\gamma \rightarrow - S_i^\gamma$, during the reproduction process. 

For simplicity, an individual is removed from the system with a constant probability $p_{kill}$ per time step (we use $p_{kill}=0.2$). This procedure is implemented both for asexual and sexually reproducing individuals. 

A time step consists of {\em one} annihilation attempt followed by {\em one} reproduction attempt. One generation consists of $N(t)/p_{kill}$ time steps, which is the average time taken to kill all currently living individuals.

Initially we place $N(0)=500$ individuals at randomly chosen positions. The results are independent of initial conditions. We obtain the same results if all individuals are located at the same position initially.

The present paper's main focus is on the asexual mode of reproduction and results presented are for asexual individuals except otherwise stated.  For completeness and for comparison we do, however, in Secs. \ref{competition} and  \ref{para} consider sexual reproduction, simplicity which is defined as follows.

\section{\label{dyn_stab} Dynamical stability}
Neglecting fluctuations in the occupancy $n({\bf S},t)$ the above dynamics is described by the following set of equations (one equation for each position in the genotype space):
\begin{eqnarray}
&&n({\bf S},t+1) = n({\bf S},t)\nonumber\\
&+&\{p_{off}({\bf S},t)[2(1-p_{mut})^L-1]-p_{kill}\}{n({\bf S},t)\over N(t)}\nonumber\\
&+&2p_{mut}(1-p_{mut})^{L-1}\sum_{\lan {\bf S}',{\bf S}\ran}
p_{off}({\bf S}',t){n({\bf S}',t)\over N(t)},
\label{mean_field}
\end{eqnarray}      
where the sum is over the nearest neighbors of ${\bf S}$. 
Stationary solutions require the system to find configurations in genotype space for which all positions satisfy the demand that either $n({\bf S},t)=0$ or if $n({\bf S},t)\neq0$ (neglecting the mutational back flow represented by the last term in Eq. (\ref{mean_field})), we must have
\begin{equation}
p_{off}={p_{kill}\over 2(1-p_{mut})^L-1}\equiv p_{q-ESS}
\label{balance}
\end{equation}
The fitness $p_{off}({\bf S}^a,t)$ of individuals at a position   ${\bf S}^a$ depends on the occupancy $n({\bf S}^b,t)$  of all the sites  ${\bf S}^b$ with which site ${\bf S}^a$ is connected through couplings $J^{ab}$. Accordingly, a small perturbation in the occupancy at one position may be able to disturb the balance in Eq. (\ref{balance}) between  $p_{off}({\bf S},t)$, $p_{kill}$ and $p_{mut}$  on connected sites. In this way an imbalance at one site can spread as a chain reaction through the system, possibly causing a global reconfiguration of the occupancy in genotype space.

\begin{figure}
\includegraphics*[angle=-90,width=8cm]{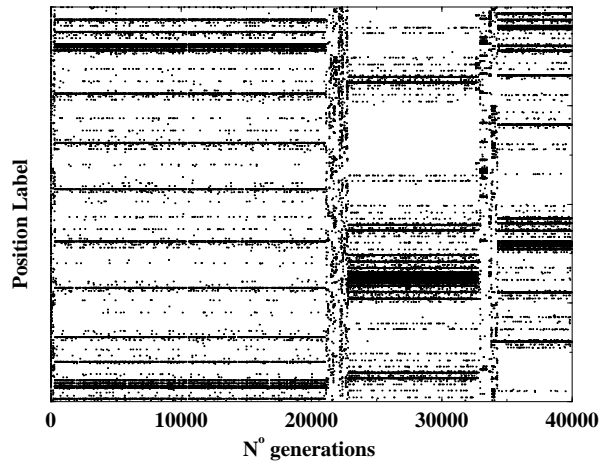}
\caption{\label{occu} The occupation in genotype space plotted as a function of generation time. The genotypes are enumerated in an arbitrary manner. If a position is occupied at a given moment in time a dot is placed at the corresponding number along the y-axis at that instant in time. Parameters are $c=0.5$, $\mu=0.005$ and $Lp_{mut}=0.25$.}
\end{figure}

We show in Fig. \ref{occu} the occupancy in genotype space plotted as a function of time for asexual reproduction. Periods of stable configurations are separated by fast transitions. We have called the stable periods ``quasi-Evolutionary Stable Strategies'' or q-ESS since they are reminiscent of the Evolutionary Stable Strategies (ESS) introduced by Maynard Smith \cite{maynard_smith}. 

It is interesting to investigate just how stable the q-ESS are. We have done this by applying different types of perturbations in the q-ESS. The result is that the q-ESS are very stable against global perturbations such as a brief or a lasting increase in control parameters $\mu$, $c$ or $p_{kill}$. Changes of up to 50\% in these parameters, either permanently or for a period of 100 generations, only effect the total population size and is typically not able to kick the population out of its present q-ESS configuration in genotype space. In contrast, a similar perturbation of the mutation rate easily destabilizes the q-ESS configuration.

We stress that the segregation (or speciation) to be discussed below is an effect of different couplings between different positions ${\bf S}^a$ and ${\bf S}^b$. When we assume $J({\bf S}^a,{\bf S}^b)=J_0$ independent of ${\bf S}^a$ and ${\bf S}^b$, the population is not concentrated around a subset of the positions in genotype space, instead the population is smeared out through the space in a diffuse manner.  Self-interaction, however, can cause segregation in a rather trivial way. Namely, if we include a distribution of $J({\bf S},{\bf S})$ values, segregation may occur even in the case where all interaction terms assume the same value: 
$J({\bf S}^a,{\bf S}^b)= J_0$ for ${\bf S}^a\neq {\bf S}^b$. However, this type of selection of configurations in genotype space is not very interesting since the sites to become occupied is determined by the arbitrarily assigned self-interactions $J({\bf S},{\bf S})$ and not by the collective dynamical adaptation at play when $J({\bf S},{\bf S})=0$ and $J({\bf S}^a,{\bf S}^b)$ assumes a distribution of different $J$-values. In reality one will expect the selection of species to be caused by a mixture of self-interaction and interaction between different species. To decide which one is dominant might be difficult and will certainly be system specific.

\begin{figure}
\includegraphics*[angle=-90,width=8cm]{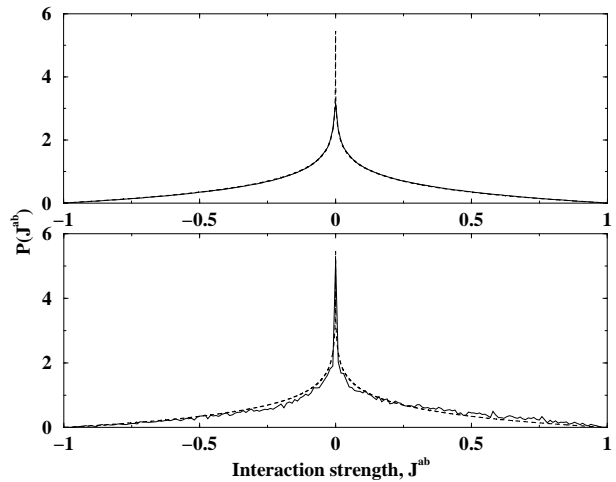}
\caption{\label{P(J)} The distribution from which the values of the couplings $J({\bf S}^a,{\bf S}^b)$ are draw at the start of the simulations (dashed curve) together with the probability density function of the couplings between occupied sites (solid curve) during the hectic periods (top panel) and during the q-ESS (bottom panel). Parameters are $c=0.01$, $\mu=0.01$ and $Lp_{mut}=0.2$.
} 
\end{figure}

\begin{figure}
\includegraphics*[angle=-90,width=8cm]{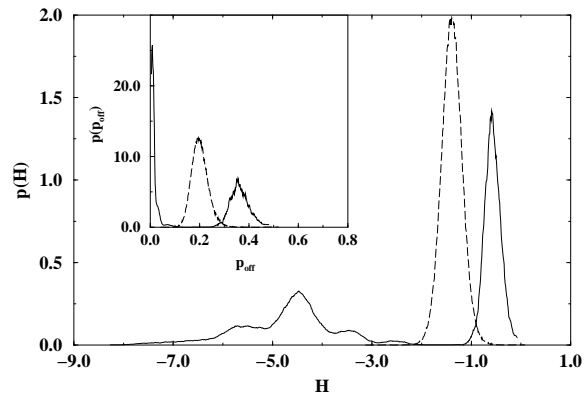}
\caption{\label{p(p_off)} The probability density function for the weight function $H$ (main frame) and reproduction rates $p_{off}$ (insert) during the hectic transitions (dashed curve)  and in the q-ESS (solid curve). Parameters are $c=0.08$, $\mu=0.005$ and $Lp_{mut}=0.25$.}
\end{figure}

There is a significant difference between the distribution of active couplings,
$p_{act}(J({\bf S}^a,{\bf S}^b))$, in the q-ESS and the distribution during the hectic transitions.
We show in Fig. \ref{P(J)} the distribution from which the  $J_{bare}({\bf S}^a,{\bf S}^b)$ are sampled together with the distribution of couplings between occupied sites after a large number of generations.  During the hectic phases there is clearly no noticeable difference between the ``bare'' distribution of the $J({\bf S}^a,{\bf S}^b)$ and the distribution of active couplings, i.e. couplings between occupied positions.  During the q-ESS we observe a slight bias towards positive $J$-values of the active couplings. This slight shift towards more positive couplings will, according to Eqs. \ref{Hamilton2} and \ref{p_off}, lead to an increased reproduction rate during the q-ESS. The manifestation of this difference between the hectic periods and the q-ESS is illustrated in Fig. \ref{p(p_off)}. We see that the distribution of $H$-values in the q-ESS during both modes of the dynamics contains a narrow peak. In the q-ESS, the peak in $p(H)$ is separated from a strongly negative band of support. The values of $H$ in this band are so negative that the corresponding $p_{off}(H)$ are negligible (see the insert in Fig. \ref{p(p_off)}). Genotype positions corresponding to this band consist of unfit positions next to highly occupied and very fit positions. The reason these positions are occupied at all is that they are supplied by mutations occurring on the neighboring fit positions. The conclusion of these considerations is that the dynamics during the q-ESS as well as during the hectic periods are controlled by the reproduction of individuals with $H$-values in the two respective peaks of $p(H)$.

The location of the peaks of $p(H)$ is determined in the following way. During the hectic periods the occupation of positions in genotype space is highly unstable and
$n({\bf S},t+1)$ is only related to $n({\bf S},t)$ in an erratic way, the balance equation Eq. (\ref{mean_field}) is never fulfilled for nonzero $n({\bf S},t)=0$.  The only constraint on $p_{off}$ during the hectic periods is accordingly that the total population remains constant on average which implies that on average  $p_{off}=p_{kill}$. This explains why in Fig. \ref{p(p_off)} the peak in $p(H)$ during the hectic periods corresponds to a peak in $p(p_{off})$ centered at $p_{kill}=0.2$. The situation is different during the q-ESS. Here the occupation of the selected positions in genotype space remains approximately constant and Eq. (\ref{balance}) applies. Substituting the relevant values $p_{kill}=0.2$, $p_{mut}=0.0125$ and $L=20$ into Eq. (\ref{balance}) produces $p_{off}=0.36$ which explains the position of the peak in $p(p_{off})$ during the q-ESS. 
  
\section{\label{aging} Aging}
For simplicity we concentrate again on the asexual model in this section.
For large genome length $L$ the system is always in a transient. The time needed to reach the stationary state increases exponentially with $L$ and is therefore unreachable for any biologically relevant values of $L$.

\subsection{Increasing q-ESS durations}
\begin{figure}
\includegraphics*[scale=.5,angle=-90,width=8cm]{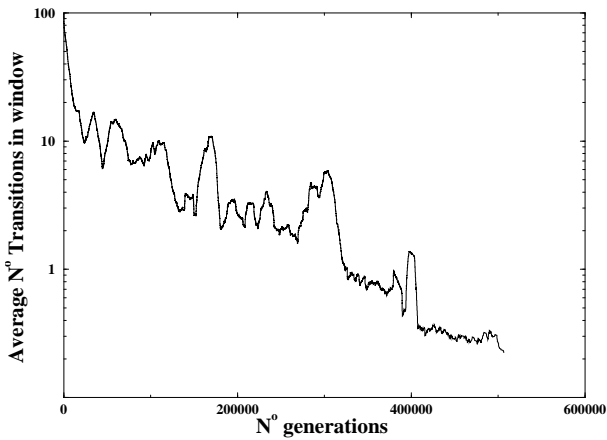}
\caption{\label{ave_N_trans} The average number of transitions
during a window of size $T=1000$ generations as a function of generation
time. Parameters are $c=0.01$, $\mu=0.01$ and $Lp_{mut}=0.2$. Tthe average is over 400 realizations.}
\end{figure}

The gradual change in the statistical measures of the model is seen directly as a slow increase with time of the average duration of the q-ESS. To demonstrate this we show in Fig. \ref{ave_N_trans} the average number of transitions $\Omega_T(t)$ between q-ESS within a time window of fixed size $T$ as a function of time
$t$ measured in number of generations. It is clear that $\Omega_T(t)$ decreases with increasing $t$, however it is very difficult to obtain sufficient statistics to be able to
determine the functional dependence of $\Omega_T(t)$ on $t$,
though a very slow exponential $t$ dependence is suggested by Fig. \ref{ave_N_trans}. Despite these sampling difficulties, it is evident that the duration of the q-ESS, on average, increases with time.
This corresponds to a decrease in the extinction rate, consistent with
analysis of the fossil record \cite{newman}.

\subsection{Increasing population size, diversity and complexity}
\begin{figure}
\includegraphics*[angle=-90,width=8cm]{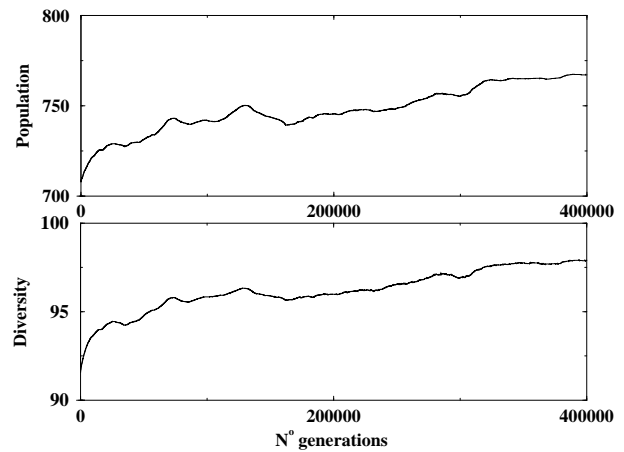}
\caption{\label{pop_div} The ensemble averaged total population
(top) and diversity (bottom) as function of generation time for the same ensemble as in Fig. \ref{ave_N_trans}. }
\end{figure}

The gradual growth of the duration of the stable q-ESS epochs indicates that the dynamics of the system is able to produce more stable or better adapted configurations in genotype space. It is difficult to test quantitatively the stability of the q-ESS with respect to perturbations. That the population is distributed in an increasingly more efficient manner in genotype space can be seen directly from the increase in the total
population size $N(t)$ averaged over an ensemble of different realizations of the stochastic elements of the dynamics.
Fig. \ref{pop_div} contains the average total population $\lan N(t)\ran$
together with the ensemble average of the diversity $\lan D(t)\ran$, where  $D(t)$ is defined as the number of different occupied positions ${\bf S}$ in genotype space at time $t$. The average diversity also increases with time.

Let us briefly consider how the total population size can increase.
We saw in Sec. \ref{dyn_stab} that essentially $p_{off}$ is
narrowly distributed either about $p_{kill}$ or, as in the q-ESS, about the value 
$p_{q-ESS}$ in Eq. (\ref{balance}). The increase in $N(t)$ is therefore not an effect of a gradual increase in $p_{off}$. Simulations indeed confirm that the average offspring probability always remains constant over the entire run. In biological observations and experiments, for reasons of uniqueness, the reproduction rate is identified as {\em fitness}. In this sense
the fitness of the individuals remains, on average, constant in the TaNa model,
as presumably is also the case in biological macroevolution; though
the microbial experiments by Lenski \cite{lenski} demonstrate that the
reproductive fitness can increase as a result of adaptation in microevolution.

The increase in the average population size $\lan N(t)\ran$
observed in the TaNa model is caused by
the system ability to generate configurations that increase the
interaction term in the weight function $H({\bf S},t)$ defined in 
Eq. (\ref{Hamilton2}). When the first term increases, the second
term $\mu N(t)$ in Eq. (\ref{Hamilton2}) can increase as well,
while the total $H({\bf S},t)$ remains on, average, fixed.

\begin{figure}
\includegraphics*[angle=-90,width=8cm]{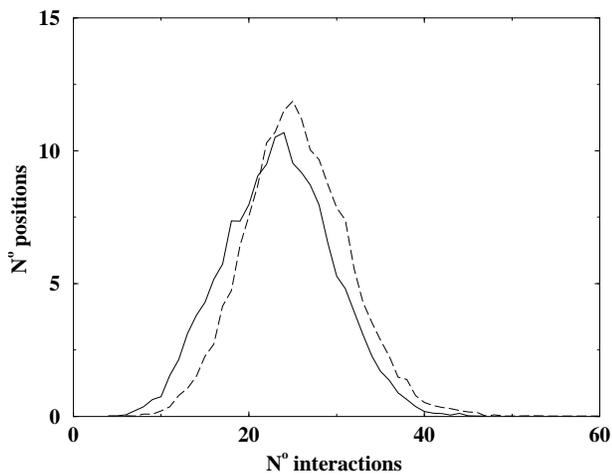}
\caption{\label{N_inter} The number of occupied positions in genotype space with
a given number of active links connected to other occupied position in genotype
space. The solid line is after $500$ generations, the dashed line
is recorded after $10^5$ generations. Parameters are $c=0.01$, $\mu=0.01$
and $Lp_{mut}=0.2$.}
\end{figure}

The increase in the interaction term of $H({\bf S},t)$ is achieved in several ways. Firstly, the population is spread out onto an increasing number $D(t)$ 
of different genotypes, as seen in Fig. \ref{pop_div}. Moreover, the evolutionary
dynamics  tends to produce occupied sites which are interacting with
an increased number of other occupied sites, i.e., the number
of non-zero terms $J({\bf S}^\alpha,{\bf S})n({\bf S})$ in 
$H({\bf S}^\alpha,t)$ in Eq. (\ref{Hamilton2}) grows as the
system produces configurations that are able to benefit better from the possible mutual 
interactions represented by $J({\bf S}^\alpha,{\bf S})$. 
The distribution of active interaction links is shown at an early and a
much later time in Fig. \ref{N_inter}. We have not been able to resolve a shift with time in the distribution of the values of the active interaction strengths $J({\bf S}^a,{\bf S}^b)$.

The increase in the diversity and the number of active links connected to an occupied 
position can be interpreted as an increase in the complexity of the configurations produced by the evolutionary dynamics. Selection and adaptation operate at the level of the entire configuration in genotype space rather than at the level of individual genotypes. 
This highlights that the biological concept of fitness makes most sense when 
considered as a {\em collective property of an ecology}, rather than an observable characteristic of the individual species or individual members of a population.

\subsection{Record Statistics}
The observation in the previous section that the dynamics of
the TaNa model leads to an increase in a number of measurable
quantities, taken together with the intermittent nature of the dynamics,
suggests that the transitions between consecutive q-ESS epochs
correspond to record transitions. One can imagine
that some characteristic measures of the collective level of 
adaptation of the configurations generated in genotype space
achieves an ever increasing value as the system undergoes a transition from one q-ESS to the next. 

Sibani and collaborators \cite{sibani_1,sibani_2,sibani_3,sibani_4,sibani_littlewood} have studied record dynamics and shown that the probability for $n$ records in a sequence of $t$ independently drawn random numbers is Poisson distributed
on a logarithmic time scale, or equivalently: that the
logarithm of the ratio of the time between the $k$-th and the 
$(k-1)$-th record, $\tau_k=\ln(t_k/t_{k-1})$, is exponentially
distributed: $P(\tau>x)=\exp(-\lambda x)$.  
Sibani and coworkers \cite{sibani_1,sibani_2,sibani_3,sibani_4} have also demonstrated the relevance of record statistics to the dynamics of the Kauffman  NK model \cite{NK-model,kauffman}.
 
\begin{figure}
\includegraphics*[angle=-90,width=8cm]{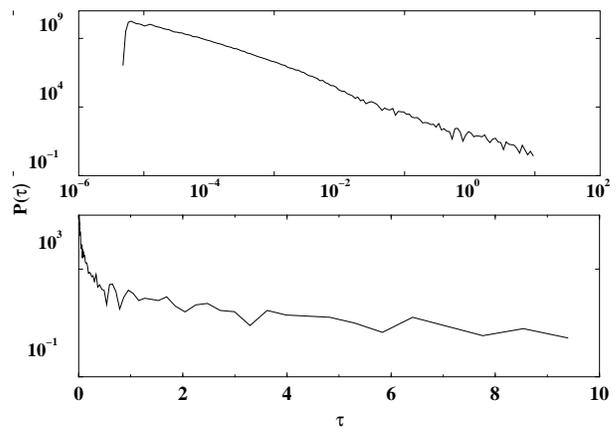}
\caption{\label{sibani} The probability density $p(\tau)$ of the logarithmic
waiting time $\tau$. The top graph is a double logarithmic
plot of $p(\tau)$ exhibiting a power law behavior in the region
of small $\tau$ values. The bottom plot is a linear-log plot of the
same data. Here one sees that the behavior of $p(\tau)$ for
large values of $\tau$ is consistent with a slow
exponential decay. Parameters are as for Fig. \ref{N_inter}.}
\end{figure}

Accordingly, it is interesting to investigate if the aging observed in the TaNa model exhibits signs of record statistics.
To do this, we study the distribution of the variable
$\tau_k=\ln (t_k/t_{k-1})$ where $t_k$ denotes the time at
which the $k$-th transition between consecutive q-ESS epochs occurs. We show in Fig. \ref{sibani} that $\tau_k$ is exponentially distributed for large values
and algebraically distributed for small values of $\tau_k$.

The exponential tail in Fig. \ref{sibani} suggests that the transition times in the TaNa model follow record statistics in the region of large $\tau$ values, corresponding to the regime of long q-ESS durations. The algebraic form of $p(\tau)$ for small $\tau$ values indicates significant correlations for transitions that occurred in rapid succession.  The question is which quantity evolves according to record dynamics. We have so far not been able to identify a variable of the system which jumps monotonously to ever higher values at the transition times. In the evolution
models studied by Sibani {\it et al.} \cite{sibani_1,sibani_2,sibani_3,sibani_4} the fitness increases through consecutive records. As mentioned above in the TaNa model the reproductive fitness $p_{off}$ remains, on average, constant. The increase in the average duration of the q-ESS (see Fig. \ref{ave_N_trans}) suggests that the stability of the configurations in genotype space gradually increases. To explore this,
one should study the temporal behavior of the eigenvalue spectrum of the
stability matrix of the effective evolution equations in (\ref{mean_field}). We expect that the number of unstable directions, on average, decreases with time, though for a given realization fluctuations probably prevent a strictly monotonous behavior.

\section{\label{spec_ab} Species abundance function}
\begin{figure}
\includegraphics*[angle=-90,width=8cm]{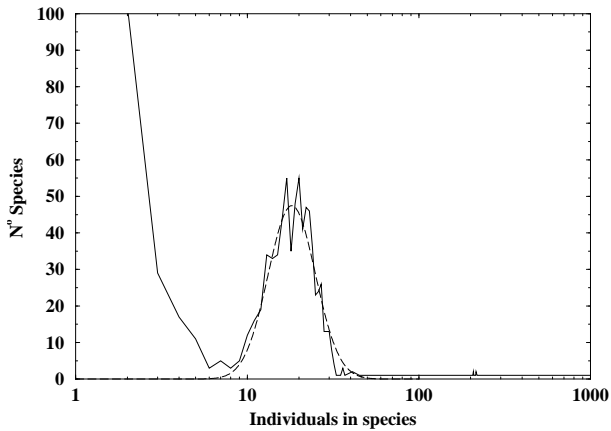}
\caption{\label{spe_ab} The species abundance distribution. The
peak in the distribution is compared with the lognormal 
form (dashed curve). Parameters are $c=0.01$, $\mu=5\cdot10^{-5}$ and $Lp_{mut}=0.2$.}
\end{figure}

The fundamental quantity to describe an ecology is the species abundance function \cite{hubbell}.  The species abundance $W(\rho)$ is the ratio $W$ of species which contains a ratio $\rho$ of the total population. 
We plot in Fig. \ref{spe_ab} the species abundance function for the TaNa model during a q-ESS, in this case we use the term species to denote individual positions in genotype space.
A large number of positions are occupied by a small number of individuals; the occupancy of these positions is never established for extended periods during the q-ESS. The robust species contain a reasonable number of individuals and are distributed according to the broad peak. The peak can be fitted by a log-normal curve in a way similar to observed species abundance functions, see e.g. \cite{hubbell}. We note that comparable species abundances functions are found in the predator-prey model studied by McKane, Alonso and Sol\'e \cite{mckane}.

\section{\label{competition} Competition between sexual and asexual reproduction}
We now consider the competition within the TaNa model between sexual and asexual reproduction in simulations of mixed populations. This is done by adding an extra {\em reproduction determining} gene to the genome (making $L=21$ in this section). This extra gene does not explicitly enter in $H({\bf S},t)$ but dictates an individual's reproductive mode. Mutations to this gene occur during reproduction in the normal way. Obviously, the simplistic nature of the TaNa model excludes the model from realistically representing all biological features of the difference between sexual and asexual reproduction. Perhaps the most essential difference between sexual and asexual reproduction is the reshuffling of genes caused by the crossing over and recombination involved in sexual reproduction \cite{burt}. It is possible that this effect can be captured by our definition of sexual reproduction for $d_{max}\geq 1$. To make the two reproductive modes as similar as possible in all aspects, except for the mixing of parent genes in the sexual case, we redefine in this section slightly the reproduction procedures in the following way. The only difference compared with the definitions in Sec. \ref{reproduction} is that an asexual reproducing individual produces {\em one} new individual and we leave the parent in the system. In sexual reproduction the reproduction probability is assumed to be $p_{off}$ of the first of the two selected parents.

According to Weismann, sexual reproduction is more efficient to adapt to changed conditions because recombination produces a larger variation of types upon which Natural Selection is able to act, see Burt's excelent review \cite{burt}. It follows that in a harsher environment, sexual reproduction will be superior to asexual; a phenomenon which is observed \cite{burt}. 
The TaNa model in its present form is not entirely able to account for these facts. It is informative, nevertheless, to investigate the behavior of the model and to understand why it fails in this particular respect. 

\begin{figure}
\includegraphics*[angle=-90,width=8cm]{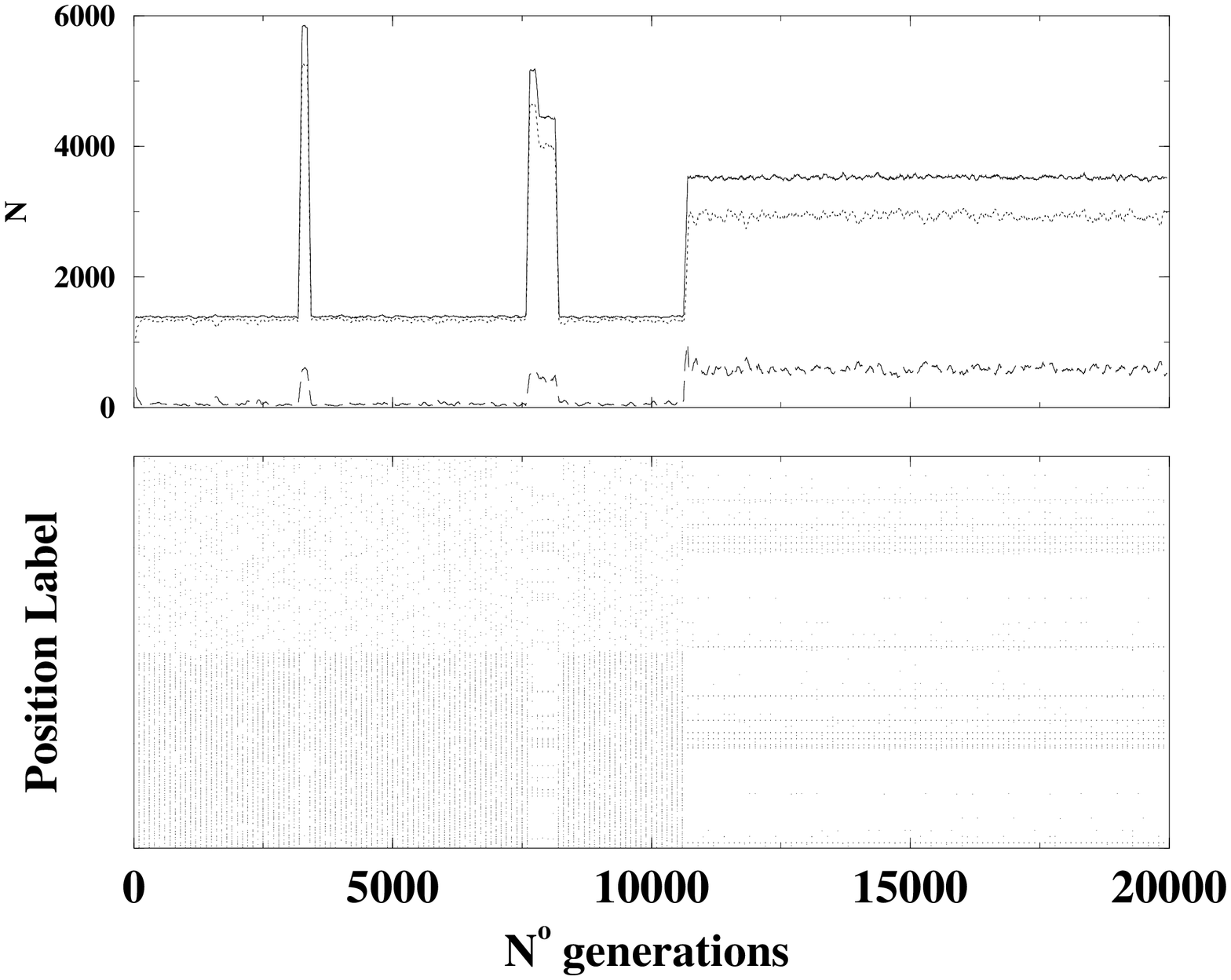}
\caption{\label{mixed_occu} Run with mixed population. Top frame contains the size of the total population (solid curve), the asexual population (dotted curve) and the sexually reproducing subpopulation (dashed curve). The frame below shows the occupation of the genotypes for the sexually reproducing subpopulation (top half) and the asexual subpopulation (bottom half). Parameters are
$c=0.08$, $\mu= 0.001$, $Lp_{mut}=0.3$ and $d_{max}=1$.} 
\end{figure}

In Fig. \ref{mixed_occu} we show the total size of the population together with the sizes of the sexual and asexually reproducing subpopulations. We see that the asexual population is always largest and even more so during the q-ESS. 

\begin{figure}
\includegraphics*[width=8cm]{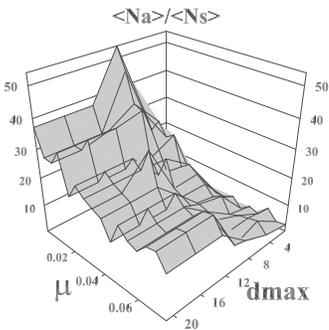}
\caption{\label{Na/Ns} The ratio of the average number of asexual
and sexually reproducing individuals as a function of 
$d_{max}$ and $\mu$. Parameters are as in Fig. \ref{mixed_occu}. } 
\end{figure}

In Fig. \ref{Na/Ns} we show the ratio of the average size $N_a$ of the asexual reproducing population and the average size $N_s$ of the sexually reproducing population. One notices that the asexual population is always more numerous, but less so for large $d_{max}$ and low $\mu$. That is the region where recombination is most important and the environment (modeled by $\mu$) least friendly. So in this respect the TaNa model exhibit a trend in the right direction in the sense of Weismann, see Burt's review \cite{burt}.

The reason the sexually reproducing population is unable to out-compete the asexual in the TaNa model is easy to understand. The offspring produced by two sexually reproducing individuals is likely to end up at a position in genotype space different from the parent positions. This effect makes it difficult to maintain the occupancy of the parent positions, even if these positions are highly fit. It is obviously this variation that is expected to make sexual reproduction more efficient in searching genotype space. However, in its present form, the TaNa model allows very large variations in the weight function $H$ in Eq. (\ref{Hamilton2}), even for positions very close by in genotype space. Whether this is reasonable or not depends on the interpretation of the genomic sequence ${\bf S}$. If we think of each ``gene'' $S_i$ as representing, in an averaged way, a section of length $1/L$ of the entire genome, then a separation by Hamming distance one is very significant for $L$ as small as 20; and it is reasonable that two positions in genotype space of this or larger separation may have very different weight functions $H$. On the other hand, this interpretation means that two individuals being different by a single Hamming distance differ by a major fraction of the entire genome and should therefore not belong to the same species and accordingly not be able to produce any (viable) offspring. We conclude that a more realistic study of the competition between sexual and asexual reproduction by use of the TaNa model should be possible if a more slowly varying weight function is used. Future studies will focus on this problem. 
 
\section{\label{error} The Error Threshold}
\begin{figure}
\includegraphics*[angle=-90,width=8cm]{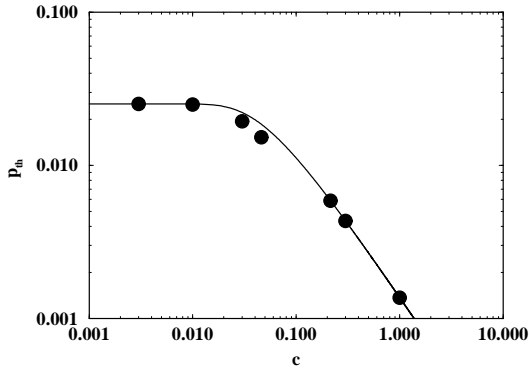}
\caption{\label{error_thres} The loss of q-ESS occurs for mutation rates
above the circles. For comparison, the theoretically predicted
error threshold $p^{th}_{mut}(c)$ is shown for $\alpha=0.07$ (see main text).  
The carrying capacity parameter is $\mu=0.005$. } 
\end{figure}

At sufficiently large mutation rates $p_{mut}$, offspring are so different from their parents that the occupation in genotype space rapidly moves from one position to the next. When this happens, it becomes impossible to establish the q-ESS seen in Fig. \ref{occu} and consequently the entire simulation consists of one hectic period. The change from the behavior depicted in Fig. \ref{occu}, where the hectic periods are of much shorter duration than the q-ESS, to the behavior where the q-ESS are absent, occurs over a very narrow region of $p_{mut}$ values. Considering first large values of  $p_{mut}$  we gradually decrease $p_{mut}$ in the simulations and we identify the threshold value, $p_{th}$ of $p_{mut}$ at which q-ESS are observed as the error threshold 
\cite{eigen,drossel_rev}. In Fig. \ref{error_thres} we plot the simulated value of $p_{th}$  for different values of the width parameter $c$.

We can estimate  the $c$ dependence of $p_{th}$ by the following argument. From Fig. \ref{p(p_off)} we know that the distribution of $p_{off}$ in the hectic periods is centered about $p_{kill}$ and in the q-ESS is centered about $p_{q-ESS}$ defined in Eq. (\ref{balance}). Changing the parameter $c$ will change the width of the distribution of the $p_{off}$ values (see Eqs. (\ref{Hamilton2}) and Eq. (\ref{p_off})).
It will be possible to establish q-ESS in between the hectic periods if $p_{kill}+\sigma_p\leq p_{q-ESS}$, where $\sigma_p$ is the half width of the peak in the distribution of $p_{off}$ in the hectic periods. We translate this argument to the distribution of the $H$ values and obtain the following estimate for $p_{th}$:
\begin{equation}
p_{th}=1-2^{-1/L}[(1-p_{kill})e^{-\alpha/c}+1+p_{kill}]^{1/L}.
\end{equation}  
Here we have assumed that the width $\sigma_p$ of the peak in the distribution of $H$ values will be given by $\sigma_p=\alpha/c$ (see Eq.(\ref{Hamilton2})) in which case $\alpha$ is a measure of the standard deviation of the factor in Eq. (\ref{Hamilton2}) multiplying $1/c$. We have used $\alpha=0.07$ to fit the simulation data in Fig. \ref{error_thres}. This value is somewhat larger but of the right order of magnitude as the corresponding quantity measured during the simulation.     

\section{\label{para} Parameter dependence}
For completeness, we present here the dependence on 
the parameters $c$ and $\mu$ in the Hamiltonian given in 
Eq. (\ref{Hamilton2}). 

\begin{figure}
\hspace*{-3cm}
\includegraphics*[width=8cm]{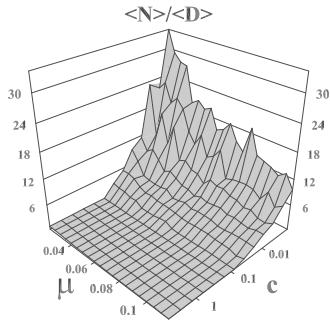}
\caption{\label{N/D_as} The ratio between the average size of
the population and the average diversity as function of the
width parameter $c$ and the physical environment parameter $\mu$.
The data are for a system with asexual reproduction with $Lp_{mut}=0.2$.} 
\end{figure}

\begin{figure}
\hspace*{-3cm}
\includegraphics*[width=8cm]{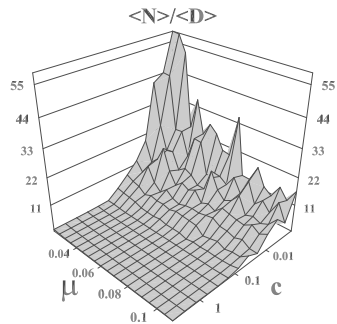}
\caption{\label{N/D_s} The same data and parameters as in Fig. \ref{N/D_as}
except that the data in this figure is for a sexually reproducing
population.} 
\end{figure}

We show in Figs. \ref{N/D_as} and 
\ref{N/D_s} the averaged occupation measured as the ratio 
between the average number of individuals and the average 
number of occupied positions in genotype space for purely 
asexual and sexual reproducing populations respectively. 
As expected, the system is able to support the largest 
populations in the region of small $\mu$ parameter and 
broad distribution of coupling strength, i.e., small values of $c$. 
The sexual reproduction is most sensitive to a decrease in the carrying capacity (increase in $\mu$) or a decrease in the width of the range of possible 
$J({\bf S}^a,{\bf S}^b)$ couplings (increase in $c$).

\section{\label{conclusion} Discussion and Conclusion}

The Tangled Nature model may be considered as a mathematical framework for 
the study of evolutionary ecology. The dynamics of the model is defined at the
level of individuals, either as asexual or as sexually reproducing individuals.
All ecological structure in the model arises through emergence.
The model is able to generate many of the observed features of
biological evolution starting from a basic implementation of 
the key assumptions in the Darwinian evolution paradigm.

The density of individuals in genotype space segregates corresponding  
to the emergence of distinct species. The interaction between individuals 
gives rise to a jerky or intermittent macro-dynamics, in which quasi-stable 
configurations (the q-ESS) in genotype space are abruptly replaced 
by new quasi-stable configurations. This mode of operation can be 
compared with the intermittent behavior observed in the fossil record and
emphasized be Gould and Eldredge's the term ``punctuated equilibrium'' \cite{gould_eldredge}. 
The TaNa model is always in a transient where the configurations generated as 
a result of adaptation to the co-evolutionary selective pressure gradually 
produce configurations, or ecologies, in genotype space which collectively 
exhibit a higher degree of adaptation, in the sense that the
average lifetime of these q-ESS increases slowly. 
This behavior compares well with the observation that the fossil record 
indicates a decrease in the extinction rate \cite{newman}.
The increase in the lifetime of the q-ESS is associated with an increase
in the complexity (species diversity, number of active interactions)  of 
successive  configurations. This gradual aging together with the
intermittent nature of the dynamics suggest that some characteristic of 
the evolving ecosystem might be undergoing record statistics in the sense 
of Sibani and co-workers \cite{sibani_1,sibani_2,sibani_3,sibani_4}. 
So far we have unfortunately not been able to identify the appropriate 
effective variable which moves through the records but we expect 
this variable to be related to the stability matrix of the effective
dynamical equation.

The species abundance distribution generated by the TaNa model 
encourage future studies of larger populations with longer genome
sequences. This will enable a hierarchical study of the
taxonomic organisation of the generated ecologies. Using distance
criteria in genotype space one can study the clustering of
individuals into species, of species into genera, etc.
Future studies will also examine the phylogenetic
structures in detail, especially during the radiation of species
encountered in the transition periods between q-ESS.
Using longer genome sequences and a more smoothly
varying weight function, we expect the TaNa model to be able
to illuminate the evolutionary competition between sexual 
and asexual reproduction.

\begin{acknowledgments}
It is a pleasure to acknowledge very helpful discussions with
Paolo Sibani. MH and SAC are supported by EPSRC studentships
and  KC gratefully acknowledges the financial support of EPSRC 
trough Grants No. GR/R44683/01 and GR/L95267/01.
We thank Paul Anderson for reading the manuscript.
 
\end{acknowledgments}

\bibliography{evolution}

\begin{thebibliography}{34}
\expandafter\ifx\csname natexlab\endcsname\relax\def\natexlab#1{#1}\fi
\expandafter\ifx\csname bibnamefont\endcsname\relax
  \def\bibnamefont#1{#1}\fi
\expandafter\ifx\csname bibfnamefont\endcsname\relax
  \def\bibfnamefont#1{#1}\fi
\expandafter\ifx\csname citenamefont\endcsname\relax
  \def\citenamefont#1{#1}\fi
\expandafter\ifx\csname url\endcsname\relax
  \def\url#1{\texttt{#1}}\fi
\expandafter\ifx\csname urlprefix\endcsname\relax\def\urlprefix{URL }\fi
\providecommand{\bibinfo}[2]{#2}
\providecommand{\eprint}[2][]{\url{#2}}

\bibitem[{\citenamefont{Newman and Sibani}(1999)}]{newman}
\bibinfo{author}{\bibfnamefont{M.~E.~J.} \bibnamefont{Newman}}
  \bibnamefont{and} \bibinfo{author}{\bibfnamefont{P.}~\bibnamefont{Sibani}},
  \bibinfo{journal}{Proc. R. Soc. Lond. B} \textbf{\bibinfo{volume}{266}},
  \bibinfo{pages}{1593} (\bibinfo{year}{1999}).

\bibitem[{\citenamefont{Eldredge}(1971)}]{eldredge}
\bibinfo{author}{\bibfnamefont{N.}~\bibnamefont{Eldredge}},
  \bibinfo{journal}{Evolution} \textbf{\bibinfo{volume}{25}},
  \bibinfo{pages}{156} (\bibinfo{year}{1971}).

\bibitem[{\citenamefont{Gould}(1977)}]{gould}
\bibinfo{author}{\bibfnamefont{S.~J.} \bibnamefont{Gould}},
  \bibinfo{journal}{Palaeobiology} \textbf{\bibinfo{volume}{3}},
  \bibinfo{pages}{135} (\bibinfo{year}{1977}).

\bibitem[{\citenamefont{Gould and Eldredge}(1977)}]{gould_eldredge}
\bibinfo{author}{\bibfnamefont{S.}~\bibnamefont{Gould}} \bibnamefont{and}
  \bibinfo{author}{\bibfnamefont{N.}~\bibnamefont{Eldredge}},
  \bibinfo{journal}{Palaeobiology} \textbf{\bibinfo{volume}{3}},
  \bibinfo{pages}{114} (\bibinfo{year}{1977}).

\bibitem[{\citenamefont{Eldredge and Gould}(1988)}]{eldredge_gould}
\bibinfo{author}{\bibfnamefont{N.}~\bibnamefont{Eldredge}} \bibnamefont{and}
  \bibinfo{author}{\bibfnamefont{S.}~\bibnamefont{Gould}},
  \bibinfo{journal}{Nature} \textbf{\bibinfo{volume}{332}},
  \bibinfo{pages}{211} (\bibinfo{year}{1988}).

\bibitem[{\citenamefont{Christensen et~al.}(2002)\citenamefont{Christensen,
  Hall, di~Collobiano, and Jensen}}]{theo_bio_p}
\bibinfo{author}{\bibfnamefont{K.}~\bibnamefont{Christensen}},
  \bibinfo{author}{\bibfnamefont{M.}~\bibnamefont{Hall}},
  \bibinfo{author}{\bibfnamefont{A.}~\bibnamefont{di~Collobiano}},
  \bibnamefont{and} \bibinfo{author}{\bibfnamefont{H.~J.} \bibnamefont{Jensen}}
  (\bibinfo{year}{2002}), \bibinfo{note}{arXiv:cond-mat/0104116,To appear in J.
  Theor. Biol.}

\bibitem[{\citenamefont{Drossel}(2001)}]{drossel_rev}
\bibinfo{author}{\bibfnamefont{B.}~\bibnamefont{Drossel}},
  \bibinfo{journal}{Adv. Phys.} \textbf{\bibinfo{volume}{50}},
  \bibinfo{pages}{209} (\bibinfo{year}{2001}).

\bibitem[{\citenamefont{Higgs and Derrida}(1992)}]{higgs1}
\bibinfo{author}{\bibfnamefont{P.~G.} \bibnamefont{Higgs}} \bibnamefont{and}
  \bibinfo{author}{\bibfnamefont{B.}~\bibnamefont{Derrida}},
  \bibinfo{journal}{J. Mol. Evolution} \textbf{\bibinfo{volume}{35}},
  \bibinfo{pages}{454} (\bibinfo{year}{1992}).

\bibitem[{\citenamefont{Eigen et~al.}(1988)\citenamefont{Eigen, McCaskill, and
  Schuster}}]{eigen}
\bibinfo{author}{\bibfnamefont{M.}~\bibnamefont{Eigen}},
  \bibinfo{author}{\bibfnamefont{J.}~\bibnamefont{McCaskill}},
  \bibnamefont{and} \bibinfo{author}{\bibfnamefont{P.}~\bibnamefont{Schuster}},
  \bibinfo{journal}{J. Phys. Chem.} \textbf{\bibinfo{volume}{92}},
  \bibinfo{pages}{6881} (\bibinfo{year}{1988}).

\bibitem[{\citenamefont{Hoster}(1993)}]{hostert}
\bibinfo{author}{\bibfnamefont{E.~E.} \bibnamefont{Hoster}},
  \bibinfo{journal}{Evolution} \textbf{\bibinfo{volume}{47}}
  (\bibinfo{year}{1993}).

\bibitem[{\citenamefont{Gravilets}(1999)}]{gavrilets}
\bibinfo{author}{\bibfnamefont{S.}~\bibnamefont{Gravilets}},
  \bibinfo{journal}{Am. Nat.} \textbf{\bibinfo{volume}{154}},
  \bibinfo{pages}{1} (\bibinfo{year}{1999}).

\bibitem[{\citenamefont{Gavrilets et~al.}(1998)\citenamefont{Gavrilets, Li, and
  Vose}}]{glv_1}
\bibinfo{author}{\bibfnamefont{S.}~\bibnamefont{Gavrilets}},
  \bibinfo{author}{\bibfnamefont{H.}~\bibnamefont{Li}}, \bibnamefont{and}
  \bibinfo{author}{\bibfnamefont{M.~D.} \bibnamefont{Vose}},
  \bibinfo{journal}{Proc. R. Soc. Lond. B} \textbf{\bibinfo{volume}{265}},
  \bibinfo{pages}{1483} (\bibinfo{year}{1998}).

\bibitem[{\citenamefont{Gavrilets et~al.}(2000)\citenamefont{Gavrilets, Li, and
  Vose}}]{glv_2}
\bibinfo{author}{\bibfnamefont{S.}~\bibnamefont{Gavrilets}},
  \bibinfo{author}{\bibfnamefont{H.}~\bibnamefont{Li}}, \bibnamefont{and}
  \bibinfo{author}{\bibfnamefont{M.~D.} \bibnamefont{Vose}},
  \bibinfo{journal}{Evolution} \textbf{\bibinfo{volume}{54}},
  \bibinfo{pages}{1126} (\bibinfo{year}{2000}).

\bibitem[{\citenamefont{Wright}(1932)}]{wright_1}
\bibinfo{author}{\bibfnamefont{S.}~\bibnamefont{Wright}},
  \bibinfo{journal}{Proc. 6. Int. Congress of Genetics}
  \textbf{\bibinfo{volume}{1}}, \bibinfo{pages}{356} (\bibinfo{year}{1932}).

\bibitem[{\citenamefont{Wright}(1988)}]{wright_2}
\bibinfo{author}{\bibfnamefont{S.}~\bibnamefont{Wright}}, \bibinfo{journal}{Am.
  Nat.} \textbf{\bibinfo{volume}{131}}, \bibinfo{pages}{115}
  (\bibinfo{year}{1988}).

\bibitem[{\citenamefont{Kauffman and Levine}(1987)}]{NK-model}
\bibinfo{author}{\bibfnamefont{S.~A.} \bibnamefont{Kauffman}} \bibnamefont{and}
  \bibinfo{author}{\bibfnamefont{S.}~\bibnamefont{Levine}},
  \bibinfo{journal}{J. Theor. Biol.} \textbf{\bibinfo{volume}{128}},
  \bibinfo{pages}{11} (\bibinfo{year}{1987}).

\bibitem[{\citenamefont{Kauffman}(1993)}]{kauffman}
\bibinfo{author}{\bibfnamefont{S.}~\bibnamefont{Kauffman}},
  \emph{\bibinfo{title}{The Origins of Order}} (\bibinfo{publisher}{Oxford
  University Press}, \bibinfo{address}{Oxford}, \bibinfo{year}{1993}).

\bibitem[{\citenamefont{Taylor and Higgs}(2000)}]{taylor}
\bibinfo{author}{\bibfnamefont{C.~F.} \bibnamefont{Taylor}} \bibnamefont{and}
  \bibinfo{author}{\bibfnamefont{P.~G.} \bibnamefont{Higgs}},
  \bibinfo{journal}{J. Theor. Biol.} \textbf{\bibinfo{volume}{203}},
  \bibinfo{pages}{419} (\bibinfo{year}{2000}).

\bibitem[{\citenamefont{Kaneko and Yomo}(2000)}]{kaneko}
\bibinfo{author}{\bibfnamefont{K.}~\bibnamefont{Kaneko}} \bibnamefont{and}
  \bibinfo{author}{\bibfnamefont{T.}~\bibnamefont{Yomo}},
  \bibinfo{journal}{Proc. R. Soc. Lond. B} \textbf{\bibinfo{volume}{267}},
  \bibinfo{pages}{2367} (\bibinfo{year}{2000}).

\bibitem[{\citenamefont{Bak and Sneppen}(1993)}]{bak_sneppen}
\bibinfo{author}{\bibfnamefont{P.}~\bibnamefont{Bak}} \bibnamefont{and}
  \bibinfo{author}{\bibfnamefont{K.}~\bibnamefont{Sneppen}},
  \bibinfo{journal}{Phys. Rev. Let.} \textbf{\bibinfo{volume}{71}},
  \bibinfo{pages}{4083} (\bibinfo{year}{1993}).

\bibitem[{\citenamefont{McKane et~al.}(2000)\citenamefont{McKane, Alonso, and
  Sol\'e}}]{mckane}
\bibinfo{author}{\bibfnamefont{A.}~\bibnamefont{McKane}},
  \bibinfo{author}{\bibfnamefont{D.}~\bibnamefont{Alonso}}, \bibnamefont{and}
  \bibinfo{author}{\bibfnamefont{R.~V.} \bibnamefont{Sol\'e}},
  \bibinfo{journal}{Phys. Rev. Let.} \textbf{\bibinfo{volume}{62}},
  \bibinfo{pages}{8466} (\bibinfo{year}{2000}).

\bibitem[{\citenamefont{Drossel et~al.}(2001)\citenamefont{Drossel, Higgs, and
  McKane}}]{drossel_2}
\bibinfo{author}{\bibfnamefont{B.}~\bibnamefont{Drossel}},
  \bibinfo{author}{\bibfnamefont{P.~G.} \bibnamefont{Higgs}}, \bibnamefont{and}
  \bibinfo{author}{\bibfnamefont{A.~J.} \bibnamefont{McKane}},
  \bibinfo{journal}{J. Theor. Biol.} \textbf{\bibinfo{volume}{208}},
  \bibinfo{pages}{91} (\bibinfo{year}{2001}).

\bibitem[{\citenamefont{May}(1972)}]{may}
\bibinfo{author}{\bibfnamefont{R.~M.} \bibnamefont{May}},
  \bibinfo{journal}{Nature} \textbf{\bibinfo{volume}{238}},
  \bibinfo{pages}{413} (\bibinfo{year}{1972}).

\bibitem[{\citenamefont{May and Anderson}(1983)}]{may_anderson}
\bibinfo{author}{\bibfnamefont{R.~M.} \bibnamefont{May}} \bibnamefont{and}
  \bibinfo{author}{\bibfnamefont{R.~M.} \bibnamefont{Anderson}},
  \bibinfo{journal}{Proc. R. Soc. Lond. B} \textbf{\bibinfo{volume}{219}},
  \bibinfo{pages}{281} (\bibinfo{year}{1983}).

\bibitem[{\citenamefont{Sibani et~al.}(1995)\citenamefont{Sibani, Schmidt, and
  m}}]{sibani_1}
\bibinfo{author}{\bibfnamefont{P.}~\bibnamefont{Sibani}},
  \bibinfo{author}{\bibfnamefont{M.~R.} \bibnamefont{Schmidt}},
  \bibnamefont{and} \bibinfo{author}{\bibfnamefont{P.~A.} \bibnamefont{m}},
  \bibinfo{journal}{Phys. Rev. Let.} \textbf{\bibinfo{volume}{75}},
  \bibinfo{pages}{2055} (\bibinfo{year}{1995}).

\bibitem[{\citenamefont{Sibani}(1997)}]{sibani_2}
\bibinfo{author}{\bibfnamefont{P.}~\bibnamefont{Sibani}},
  \bibinfo{journal}{Phys. Rev. Let.} \textbf{\bibinfo{volume}{79}},
  \bibinfo{pages}{1413} (\bibinfo{year}{1997}).

\bibitem[{\citenamefont{Sibani and Brandt}(1998)}]{sibani_3}
\bibinfo{author}{\bibfnamefont{P.}~\bibnamefont{Sibani}} \bibnamefont{and}
  \bibinfo{author}{\bibfnamefont{M.}~\bibnamefont{Brandt}},
  \bibinfo{journal}{Int. J. Mod. Phys. B} \textbf{\bibinfo{volume}{12}},
  \bibinfo{pages}{361} (\bibinfo{year}{1998}).

\bibitem[{\citenamefont{Sibani and Pedersen}(1999)}]{sibani_4}
\bibinfo{author}{\bibfnamefont{P.}~\bibnamefont{Sibani}} \bibnamefont{and}
  \bibinfo{author}{\bibfnamefont{A.}~\bibnamefont{Pedersen}},
  \bibinfo{journal}{Europhys. Let.} \textbf{\bibinfo{volume}{48}},
  \bibinfo{pages}{346} (\bibinfo{year}{1999}).

\bibitem[{\citenamefont{Sibani and Littlewood}(1993)}]{sibani_littlewood}
\bibinfo{author}{\bibfnamefont{P.}~\bibnamefont{Sibani}} \bibnamefont{and}
  \bibinfo{author}{\bibfnamefont{P.}~\bibnamefont{Littlewood}},
  \bibinfo{journal}{Phys. Rev. Let.} \textbf{\bibinfo{volume}{71}},
  \bibinfo{pages}{1482} (\bibinfo{year}{1993}).

\bibitem[{\citenamefont{Wagner et~al.}(1998)\citenamefont{Wagner, Baake, and
  Gerische}}]{wagner}
\bibinfo{author}{\bibfnamefont{H.}~\bibnamefont{Wagner}},
  \bibinfo{author}{\bibfnamefont{E.}~\bibnamefont{Baake}}, \bibnamefont{and}
  \bibinfo{author}{\bibfnamefont{T.}~\bibnamefont{Gerische}},
  \bibinfo{journal}{J. Stat. Phys.} \textbf{\bibinfo{volume}{92}},
  \bibinfo{pages}{1017} (\bibinfo{year}{1998}).

\bibitem[{\citenamefont{{Maynard Smith}}(1982)}]{maynard_smith}
\bibinfo{author}{\bibfnamefont{J.}~\bibnamefont{{Maynard Smith}}},
  \emph{\bibinfo{title}{Evolution and the theory of games}}
  (\bibinfo{publisher}{Cambridge University Press}, \bibinfo{year}{1982}).

\bibitem[{\citenamefont{Lenski and Travisano}(1994)}]{lenski}
\bibinfo{author}{\bibfnamefont{R.~E.} \bibnamefont{Lenski}} \bibnamefont{and}
  \bibinfo{author}{\bibfnamefont{M.}~\bibnamefont{Travisano}},
  \bibinfo{journal}{Proc. Natl. Acad. Sci. USA} p. \bibinfo{pages}{6808}
  (\bibinfo{year}{1994}).

\bibitem[{\citenamefont{Hubbell}(2001)}]{hubbell}
\bibinfo{author}{\bibfnamefont{S.~P.} \bibnamefont{Hubbell}},
  \emph{\bibinfo{title}{The Unified Neutral Theory of Biodiversity and
  Biogeography}} (\bibinfo{publisher}{Princeton University Press},
  \bibinfo{address}{Princeton}, \bibinfo{year}{2001}).

\bibitem[{\citenamefont{Burt}(2000)}]{burt}
\bibinfo{author}{\bibfnamefont{A.}~\bibnamefont{Burt}},
  \bibinfo{journal}{Evolution} \textbf{\bibinfo{volume}{54}},
  \bibinfo{pages}{337} (\bibinfo{year}{2000}).

\end{thebibliography}

\end{document}